\documentclass[prl,aps,twocolumn,showpacs,floatfix,superscriptaddress]{revtex4-1}

\usepackage{graphicx}
\usepackage{textcomp}
\usepackage{amsmath, amssymb}
\usepackage[usenames]{color}

\begin{document}

\title{Sink vs. tilt penetration into shaken dry granular matter: \\
the role of foundation}

\author{G. S\'{a}nchez-Colina}
\affiliation{Group of Complex Systems and Statistical Physics,
Physics Faculty, University of Havana, 10400 Havana, Cuba}

\author{A. J. Batista-Leyva}
\affiliation{Instituto Superior de Tecnolog ́ıas y Ciencias Aplicadas (InSTEC).
Salvador Allende esq. Luaces s/n. Plaza de la Revoluci\'{o}n
CP 10400. POB 6163. La Habana, Cuba and Group of Complex Systems
and Statistical Physics, Physics Faculty, University of Havana,
10400 Havana, Cuba}

\author{C. Cl\'{e}ment}
\affiliation{Institut de Physique du Globe de Strasbourg (IPGS),
Ecole et Observatoire des Sciences de la Terre (EOST), University of
Strasbourg/CNRS, 67084 Strasbourg, France}

\author{E. Altshuler}
\affiliation{Group of Complex Systems and Statistical Physics,
Physics Faculty, University of Havana, 10400 Havana, Cuba}

\author{R. Toussaint}
\affiliation{Institut de Physique du Globe de Strasbourg (IPGS),
Ecole et Observatoire des Sciences de la Terre (EOST), University of
Strasbourg/CNRS, 67084 Strasbourg, France}

\date{ \today}

\begin{abstract}

We study the behavior of cylindrical objects as they sink into a dry
granular bed fluidized due to lateral oscillations, in order to shed
light on human constructions and other objects. Somewhat unexpectedly, we
have found that, within a large range of lateral shaking powers,
cylinders with flat bottoms sink vertically, while those with a
\textquotedblleft foundation\textquotedblright consisting in a
shallow ring attached to their bottom, tilt besides sinking. The
latter scenario seems to dominate independently from the nature of
the foundation when strong enough lateral vibrations are applied. We
are able to reproduce the observed behavior by quasi-2D numerical
simulations, and the vertical sink dynamics with the help of a
Newtonian equation of motion for the intruder.

\end{abstract}

\maketitle

\section{Introduction}

\begin{figure}
\includegraphics[width=0.4\textwidth]{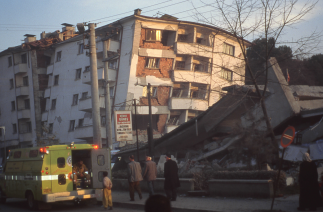}
\centering \caption{Tilted building after the Izmit Earthquake, Aug 17th,
1999, Adapazari village, Turkey. Picture: Courtesy of Mustapha Meghraoui, IPGS.}
\label{Fig1}
\end{figure}

The Kocalei earthquake occurring on August 17, 1999 affected in
various ways many constructions in the city of Adapazari, Turkey.
Following observers, some buildings sank vertically into the soil,
others tilted as shown on Fig. \ref{Fig1}, and some even suffered
lateral translation over the ground \cite{Bray1999,Sancio2002,Sancio2004}.
This case illustrates well the diversity of damage that earthquake
fluidization of soils may cause to man-made structures \cite{Ambraseys1998}.

Liquefaction in the ground may be originated dynamically, by shear
waves released during earthquakes, generating cyclic shear stresses that lead to the gradual
build-up of pore water pressure. The shaking produced by seismic events is a trigger for
extensive liquefaction, as was observed recently in Belgium \cite{Vannste1999}.

Ground fluidization \cite{NRC1985,Berril1985} has been investigated
in different kinds of media like sand \cite{Berril1985}, dry
granular soils \cite{Clement1991} and sediments \cite{Wang2014}. Of
immediate interest for engineering and for the geosciences is to
understand how man-made structures such as buildings, and massive
rocks laying on granular soils respond to fluidization associated to
seismic waves.

\begin{figure}
\includegraphics[width=0.4\textwidth]{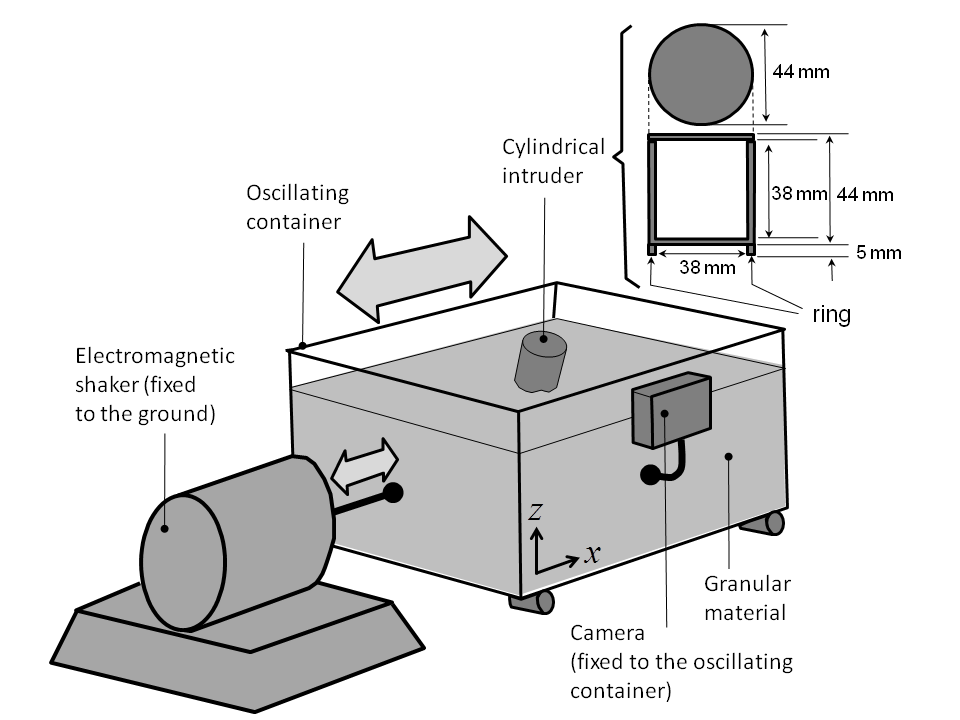}
\centering \caption{Experimental setup. At the upper right, we have
illustrated the intruder consisting in a cylinder with ring.}
\label{Fig2}
\end{figure}

Granular matter itself displays a variety of puzzling phenomena
\cite{Jaeger1996,Shinbrot1998,Altshuler2003,Aranson2006,Andreotti2013,Altshuler2008,
Boudet2007,Vinningland2007,Johnsen2006,Johnsen2008,Niebling2010a,Niebling2010b,Niebling2012,
Turkaya2015},
but during the last decade or so, our understanding of the dynamics
of objects penetrating into granular media has advanced quickly
\cite{Uehara2003,Boudet2006,Katsuragi2007,Goldman2008,Pacheco2011,Kondic2011,Torres2012,Clark2013,
Ruiz-Suarez2013,Brzinski2013,Altshuler2014,Joubaud2014,Harich2011, Kolb2004}. While
laterally shaken granular beds have received a certain degree of
attention \cite{Metcalfe2002,Kruelle2009}, the performance of
objects initially laying on the surface of a granular bed submitted
to lateral shaking has been rarely studied
\cite{Tennakoon1999,Liu1997,Sanchez2014}.

In this paper we perform systematic experiments associated to the
latter scenario, which may help understanding the performance of
human constructions and rocks  laying on granular beds during
earthquakes. In particular, using a cylinder as a simplified model
for buildings or rocks, we study its settling dynamics on a granular
bed submitted to lateral vibrations. Somewhat unexpectedly, we have
found that, within a large range of lateral shaking powers,
cylinders with flat bottoms sink vertically, while those with a
\textquotedblleft foundation\textquotedblright  consisting in a
shallow ring attached to their bottom, tilt besides sinking. The
latter scenario seems to dominate independently from the nature of
the foundation when strong enough lateral vibrations are applied.
Quasi-2D simulations were also performed mimicking the experiments.
The settling dynamics of the simulated intruders, with or without
foundation reproduces quite well the corresponding experimental
results. We also developed a model that reproduces well the sinking
dynamics and gives a qualitative explanation of the tilting process.

\section{Experimental}

The penetration experiments were performed on a granular bed
contained into a test cell of approximately $25 \times 25 \times
25$~cm$^3$ filled with Ugelstad spheres of non expanded polystyrene
with a density 1.05 g/cm$^3$, and diameter 140 $\mu$m (monodisperse
within a $1$ percent), type Dynoseeds \textcopyright , produced by
Microbeads, Norway. The box was horizontally shaken at different
amplitudes ($A$), and a frequency ($f$) of  $5.0$ Hz (a value
commonly found in seismic waves), using a TIRA TV51120 shaker, type
S51120, see figure \ref{Fig2}. The maximum amplitude of oscillations
allowed by the shaker corresponds to an acceleration of $\approx
12.2$ m/s$^2$.

Two types of intruders were used in the experiments: (a) a hollow 3D printed
cylinder of $44$~mm diameter, $44$~mm height (external dimensions), and $5$~mm
thick walls, and (b) the same cylinder with a ring of $5$~mm height and
$3$~mm thickness glued to its bottom (illustrated in the upper right
corner of Fig. \ref{Fig2}). Intruders (a) and (b) will be called
\textquotedblleft No-ring\textquotedblright and \textquotedblleft Ring\textquotedblright,
respectively, from now on. Their masses were adjusted with ballast in such a
way that their densities matched the average effective density of the granular
medium, which was measured as $0.43$ g/cm$^3$. As far as the ballast used has a
density near the effective density of the granular material, it was
almost evenly distributed inside the cylinder. Note that, using a flat
bottom cylinder and a ring-like bottom cylinder, we are modifying the
\textquotedblleft foundation\textquotedblright of our intruder.

A digital camera {\it Hero 2} made by GoPro was fixed to the
electromagnetic shaker, in such a way that it could take a video of
the sinking process from an oscillating reference frame locked to
the test cell, as proposed in \cite{Sanchez2014}. This method
allowed a much better quality of the cylinder's images, and made
easier their digital processing. Videos were taken at a maximum rate
of $120$ frames per second, with a resolution of $1920 \times 1080$
pixels.

\begin{figure}
\includegraphics[width=0.45\textwidth]{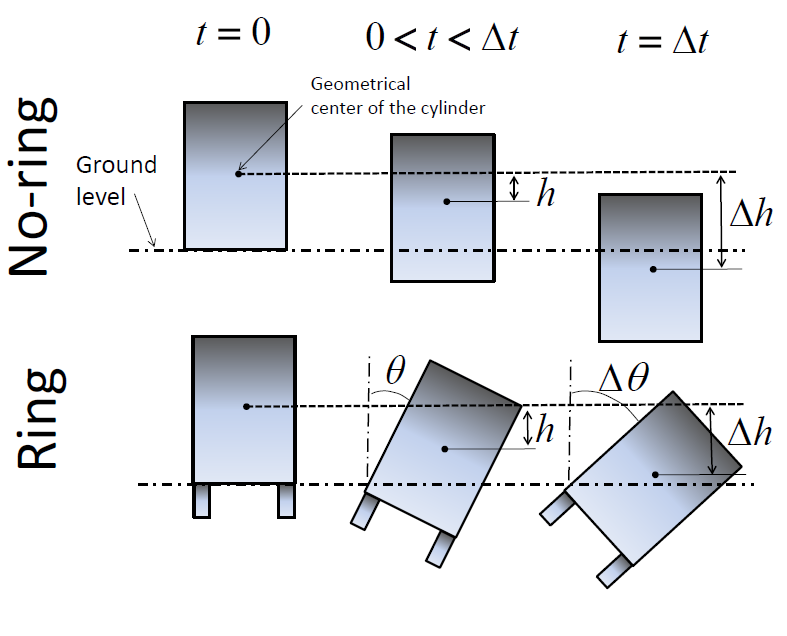}
\centering \caption{Sketch of sinking and tilting processes. The top row illustrates the
sinking process of a No-Ring cylinder in three moments during the
experimental interval from $t=0$ to a final time  $t=\Delta t$. The bottom row shows the
same temporal sequence for a Ring cylinder, which tilts in addition to
sink.}
\label{Fig3}
\end{figure}

The images were processed as follows. We first converted the videos to
image sequences in *.jpg format, and cropped each picture, excluding
irrelevant space. Then, the images were binarized through an appropriate
threshold. Using the tool {\it regionprops} from {\it MatlabR2014a}, we
identified and assigned coordinates to several bright marks we had glued
to certain points of the cylindrical intruder. The coordinates of the
marks where used to calculate the position of the intruder's geometrical
center and inclination relative to the vertical in each picture. In some
experiments where the sinking was particularly big, it was difficult to
obtain the tilt angle, since part of the marks sank below the level of the
sand surface, and they were impossible to follow. In such cases the upper
border of the cylinder was identified using the Matlab's tools {\it find} and
{\it bwtraceboundary}, and then fitted to a polynomial using the function
{\it polyfit}. The fit was used to find the inclination. In the case of
experiments ending in a very inclined position, the reference to calculate
the inclination was the cylinder's corner above the sand surface, that was
identified as the intersection of the two polynomial fits of the upper and
one lateral borders of the cylinder.

As the cylinder vibrates due to the vibration of the box, it is difficult
to determine the final position, particularly when there is a big tilting.
Then, in order to determine the sinking depth and tilting, we observe in the
videos the onset of a cyclic  movement of a reference point in the cylinder.
Once this situation was reached, the final position could be measured in the
frames filmed after the shaker was stopped.

The experimental protocol  can be described as follows: (I) preparing
the granular medium by stirring it evenly with a long rod, (II)
gently depositing the cylinder in the upright position on the free
surface of the granular bed, (III) turning ON the camera, (IV)
switching ON the shaker after setting the desired frequency and
amplitude (V) turning OFF the shaker and the camera after the
penetration process had visibly finished.

\begin{figure}
\includegraphics[width=0.45\textwidth]{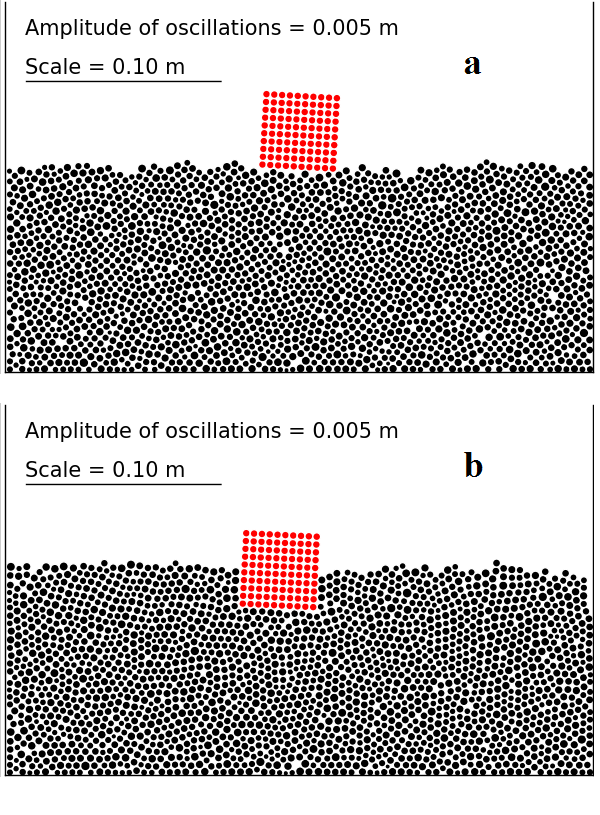}
\centering \caption{(color online) Initial (a) and final (b)
positions of No-ring intruder in a typical quasi-2D simulation using
a frequency of 5 Hz (amplitude included in the graph).}
\label{Fig4}
\end{figure}

\begin{figure}
\includegraphics[width=0.47\textwidth]{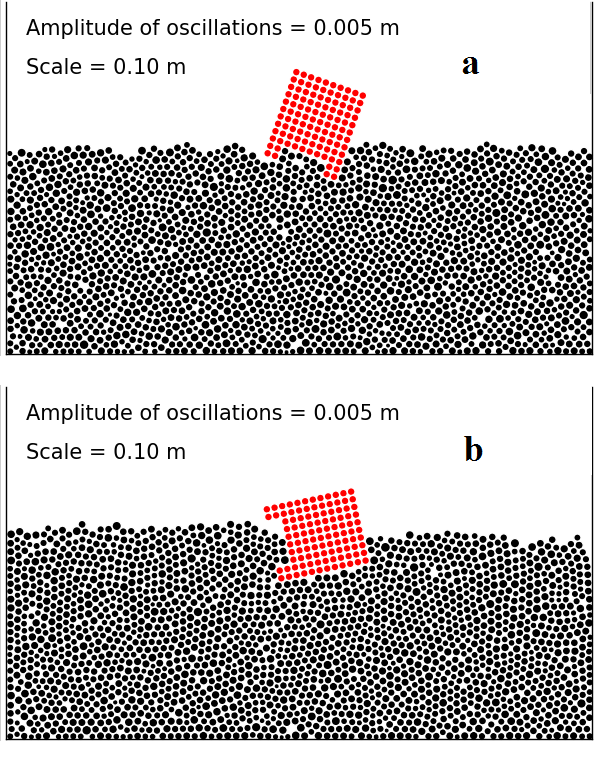}
\centering \caption{(color online) Initial (a) and final (b)
positions of Ring intruder in a typical quasi-2D simulation using a
frequency of 5 Hz (amplitude included in the graph). }
\label{Fig5}
\end{figure}

In Fig. \ref{Fig3} we define the main parameters describing the sinking process of
a No-ring cylinder (upper row), and the tilting and sinking of a Ring
cylinder (bottom row), during the experimental lapse, defined as $\Delta t$. As
 the figure indicates, in the following we will call $h$ the penetration of the
 geometrical center at a time $t$ and $\Delta h$ the final penetration at time $t=\Delta t$. It is important to
 note that both magnitudes are defined as the vertical displacement of the geometrical
 center of the cylinder (without taking into account the ring).

We also explored the observed phenomenology through numerical
simulations. They were based on a discrete element method code (DEM)
for the computation of granular systems \cite{Cundall1979,Parez2016,Johnsen2006,
Niebling2010a,Niebling2012}. We
modeled a quasi-2D granular medium, made of spheres of diameter 4 mm
monodisperse within a 1 percent to avoid the effect of
crystallization, and a thickness of 0.2 mm. The numerical medium
contains 2000 particles, filling a virtual space of 30 cm width and
10 cm height. The code calculates the position and rotating angle of
each sphere derived from the different forces applied on it. The
friction coefficient $\mu$ is taken as 0.3. To approach the
experimental conditions, we simulate particles of density 1.05
g/cm$^3$.

We created two intruders made of cohesive particles. One is a square
of 40 mm side, made of 100 particles placed in a quasi-2D square
arrangement, which simulates the No-ring intruder of the
experiments. The second one is another square of 40 mm side attached
to two small feet made of 4 particles each, mimicking the cross
section of the ring attached to the bottom of the intruder. The
particles density of the spheres which form the intruders is 1
g/cm$^3$, and the porosity of the intruders is 0.21 percent, so the
intruder density is 0.78 g/cm$^3$, i.e., approximately the same
effective density of the quasi-2D granular medium.

Once our granular medium is created, we place the intruder 1 mm
above the medium. We let it drop and settle until the whole
medium reaches equilibrium. Then, we apply a horizontal oscillation
of different amplitudes and a frequency of 5 Hz to the walls of the
medium and compute the time evolution of the position and tilting
angle of the intruder.

Figures \ref{Fig4} and \ref{Fig5} show the initial and final positions
of both types of intruders in two typical runs. Fig. \ref{Fig4} indicates
that the No-ring cylinders almost do not tilt, while in Fig. \ref{Fig5}
is obvious the wide inclination of a Ring one.

\section{Results and discussion}

\subsection{Sink {\it vs.} tilt penetration}
\begin{figure}
\includegraphics[width=0.40\textwidth]{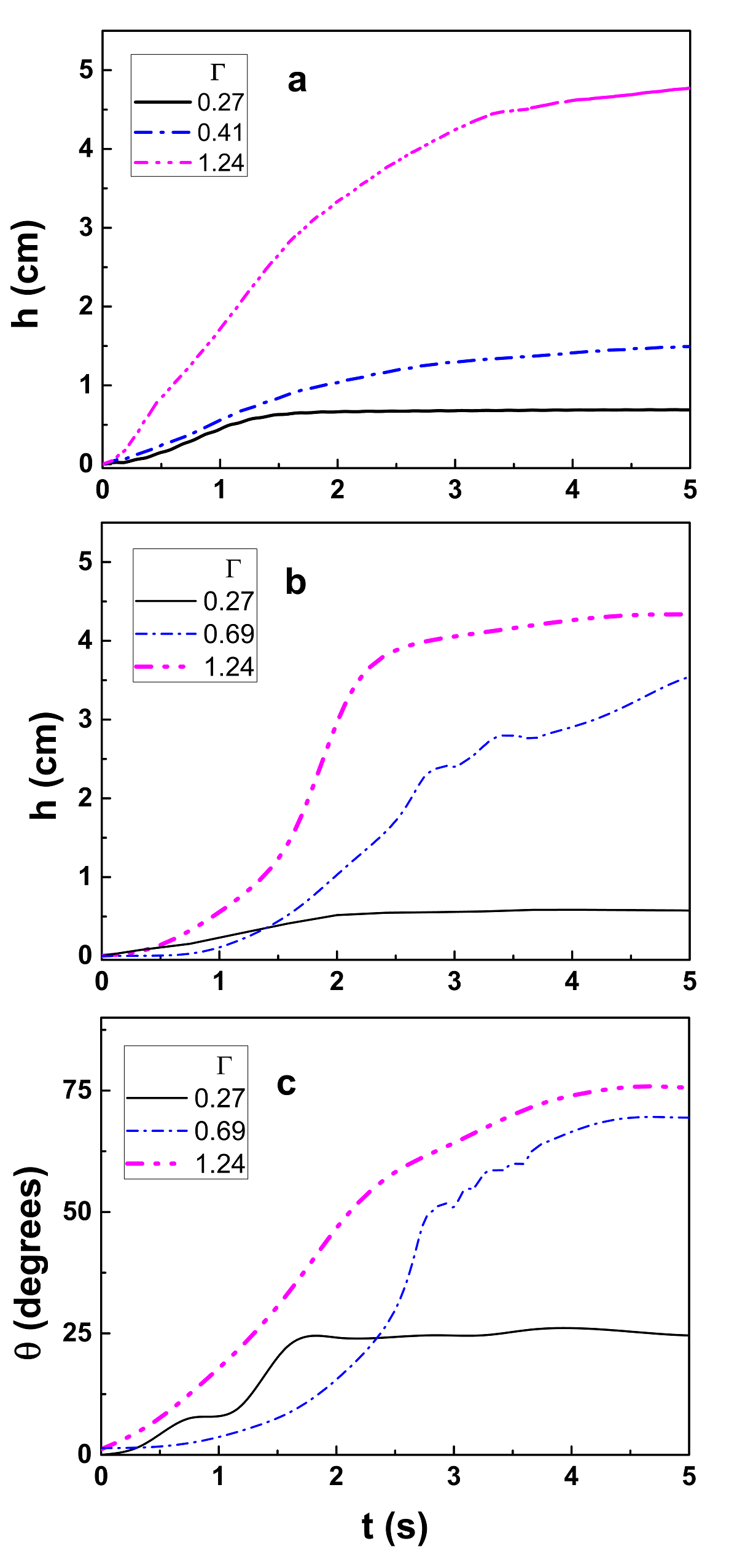}
\centering \caption{(color online) Time evolution of penetration
depths and tilt angles. Time dependence of the penetration depth of
a No-ring cylinder (a), the penetration depth of a Ring cylinder (b)
and the tilting angle of a Ring cylinder (c), for different
adimensional accelerations. The long-time creep process is not
completely shown. The tilting angle of No-ring cylinders is not
displayed, due to the fact that it oscillates around angles not
larger that $5\textrm{\textdegree}$ relative to the vertical direction}
\label{Fig6}
\end{figure}

Figure \ref{Fig6}(a) shows the time variation of the sinking depth
for selected values of the adimensional acceleration $\Gamma = A (2
\pi f)^2 / g$ (where $g=9.81$ m/s$^2$ is the gravitational
acceleration and $ A (2 \pi f)^2 $ is the horizontal peak
acceleration of the sand box) for No-ring cylinders. It is easy to
see that the penetration of the No-ring cylinders follows a common
pattern for all the accelerations. A first process of fast sinking
is followed by a slow creep. Only the penetration depth increases
with $\Gamma$. In this figure we do not show the total creep
process, due to its long duration. As the height of the cylinder is 44
mm, it is possible to check from Fig. \ref{Fig6}(a) that, for an
adimensional accelerations of 1.24, the cylinder sinks completely.

Figure \ref{Fig6}(b) is similar to the previous one, only measurements
were performed with Ring cylinders.

The general features of both graphics are very similar, but there is
a difference, that will be better observed in the following figures:
the adimensional acceleration at which the cylinder sinks completely
in the medium is bigger for the Ring cylinders than for the No-ring
ones. We could have a glimpse of why it occurs if we analyze Fig.
\ref{Fig6}(c), that presents the time evolution of the tilting angle
for a Ring cylinder. Let us analyze, for instance, the curve for the
highest adimensional acceleration shown in the figure
($\Gamma=1.24$). In the first second the inclination angle reaches a
value around 25\textdegree, resulting in the increase of the
effective size of the intruder, with the corresponding increase of
the forces impeding the sinking process (we will further develop
these ideas below). All in all, it explains why the Ring cylinder
sinks less than the No-ring one.

Another important difference is that the sinking and tilting
dynamics of Ring cylinders is more irregular than that of the
No-ring ones. It is easy to understand, if we imagine the tumbling
process. Firstly the intruder gets an initial inclination along one
of the two possible directions, due to the horizontal acceleration
provoked by the shaker, that breaks the symmetry in a non
predictable way. When the inclination appears, the gravity produces
an additional torque that increases it. But, as the granular medium
oscillates, the torque applied on the cylinder by the granular
medium changes it orientation, provoking an oscillation in the
emerged part of the intruder, so the tilting angle and the height of
the center of mass relative to the surface of the granular medium
also oscillate. This is illustrated in Fig. \ref{Fig6}(b) and (c),
even after being submitted to an averaging process.

\begin{figure}
\includegraphics[width=0.40\textwidth]{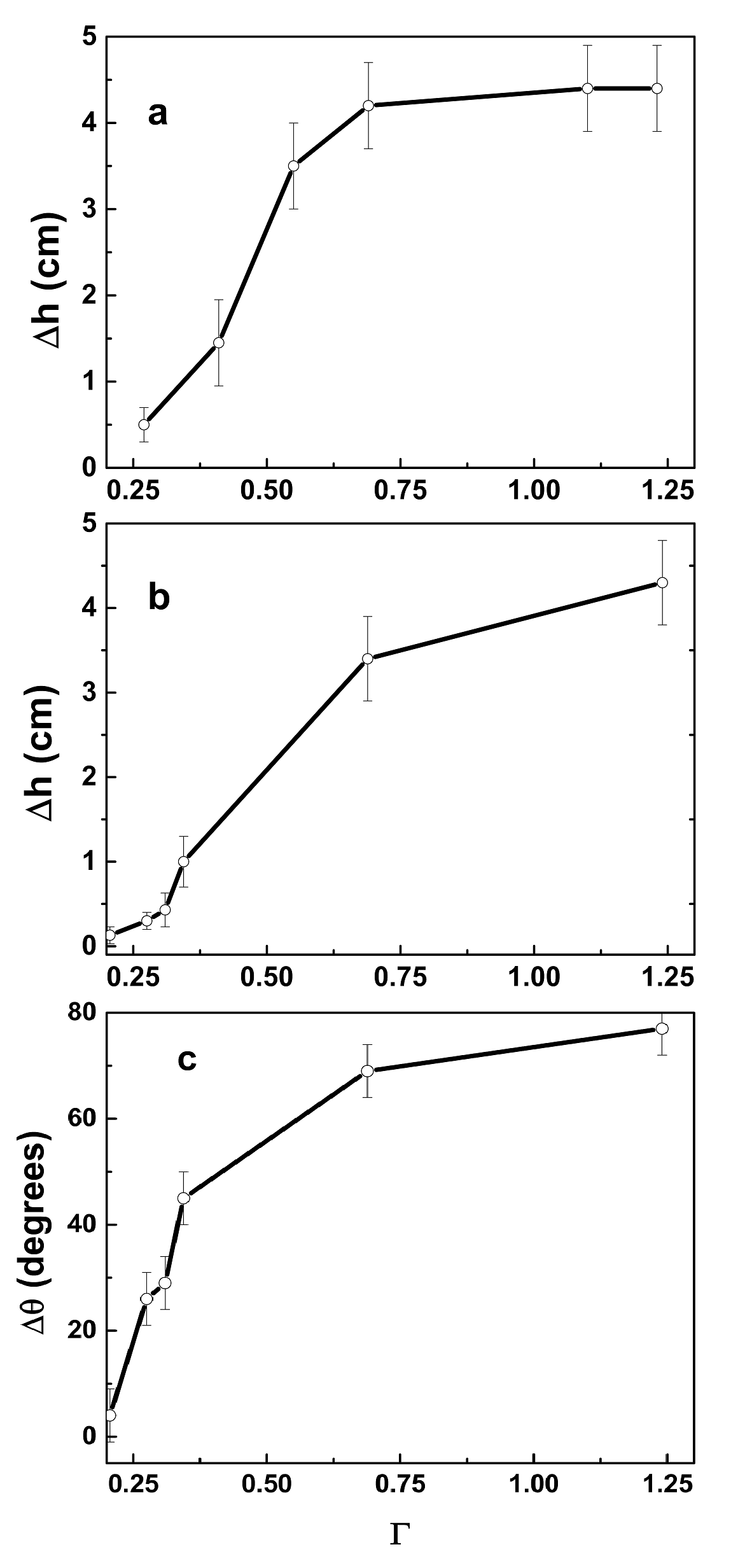}
\centering \caption{Sinking and tilting: heights and angles. Final
vs. initial sink heights for No-ring cylinders (a) and for Ring
cylinders (b). Tilt angles for Ring cylinders (c). Tilt angles of
No-ring cylinders are not shown for the same reasons of the previous
figure.} \label{Fig7}
\end{figure}

In general, No-ring cylinders tend to sink vertically as the granular soil is
fluidized by horizontal shaking, while cylinders with rings tend to tilt. Figure
\ref{Fig7} quantifies the differences between the initial and final stages of the
process, for almost all the range of accelerations our experimental setup was able
to reach.

Figure \ref{Fig7}(a) shows sink data for No-ring cylinders. As can
be seen, for adimensional accelerations up to $\Gamma$ = 0.27, there
was no significant penetration of the intruder into the granular
bed. Vertical penetrations started to increase significantly above
$\Gamma \approx 0.3$, reaching a plateau around $\Gamma \approx
0.75$. At the plateau, the cylinder has sank completely, but stays
\textquotedblleft floating\textquotedblright into the fluidized
granular medium, as expected for an object isodense relative to it.

In Fig. \ref{Fig7}(b) the sinking process of the Ring cylinders is
summarized. Differently from the previous case, the plateau is not
observed for the range of accelerations recorded. Note that as far
as the depth reached at  $\Gamma \approx$ 1.2 is approximately 4.4
cm (the height of the cylinder), for bigger accelerations this would
also be the final penetration, so we can safely imply that the
plateau appears at higher accelerations for that type of cylinders.
But even at the highest accelerations, there is always a small part
of the cylinder above the ground level.

Figure \ref{Fig7}(c) shows the tilt data for Ring cylinders.  No
significant tilting is observed for $\Gamma$ smaller than
approximately 0.25. With the increase of the adimensional
accelerations, the cylinder significantly tilts, increasing abruptly
the tilting angle with $\Gamma$, until saturation starts at $\Gamma
\approx 0.75$. As was stated in the figure caption, we do not show
the tilting angle of No-ring cylinders, because it is always smaller
than 5\textdegree, with a random distribution of values around the
vertical direction.

Figures \ref{Fig7}(b) and (c) are closely related, because they are
two descriptions of the same process: the motion of Ring cylinders
into the granular medium, that includes both sinking and tilting.
The fact that at the accelerations shown in this figure the plateau
in the sinking depth is barely reached while for the tilting angle
it is, could be explained by the increase of the friction of the
intruder with the granular medium when the tilting angle increases.
Then, at $\Gamma \approx 0.75$ the intruder has approximately
reached its maximum inclination, but is not completely submerged in
the medium. An increase in the acceleration does not increases
significantly the angle, because the resulting torque has diminished
due to the influence of both sinking and tilting, but the increase
in fluidization helps further sinking, until most of the cylinder is
submerged, reaching the plateau.

The overall behavior in Fig. \ref{Fig7}(a) can be understood taking
into account the experimental results in Ref. \cite{Tennakoon1999}. When the
system is submitted to lateral shaking, a \textquotedblleft solid\textquotedblright \
layer is formed, starting at a depth $h_f$  that depends
 on the adimensional acceleration $\Gamma$. For accelerations in the range we used,
 $h_f$ varies almost linearly with $\Gamma$ (see Fig. 3(a) in Ref. \cite{Tennakoon1999}),
 so we can write
 \begin{equation}
\label{eq.1}
h_f(\Gamma)= \alpha (\Gamma - \Gamma^*); \Gamma > \Gamma^*
\end{equation}
\noindent where $\Gamma^*$ is the onset of fluidization and $\alpha$ is the slope
of the linear dependence. If $\Gamma \leq \Gamma^*$ the depth of the fluidized
layer is zero.

According to reference \cite{Toussaint2014}, $\Gamma^*$ can be taken
as proportional to the friction coefficient $\mu$ between the
cylinder and the granular medium. So in this experiments we can
approximate $\mu \approx 0.3$, that is the value we use for the
simulations. They also conclude that the final depth of intrusion
depends on isostasy, and on the severity of shaking. It can be
entirely determined by isostasy, when the shaking completely unjam
the medium and suppresses the average friction around the intruder.

Then, at  low values of  $\Gamma$ the granular media is not
fluidized, and the cylinder almost does not sink (merely 5 mm at
$\Gamma=0.27$; see Fig. \ref{Fig7}(a)). At accelerations above the
fluidization threshold, the cylinder sinks until it gets in contact
with the solid layer. The larger is the acceleration, the deeper is
that layer, so the bigger is $\Delta h$. But as soon as the solid
layer appears at a depth larger than the cylinder's height, it does
not sink further: instead, it \textquotedblleft
floats\textquotedblright \ due to isodensity with the sand, so a
plateau is reached.

The strong differences in the dynamics of No-ring and Ring cylinders
within the adimensional acceleration range $0.3 \leq \Gamma \leq 1.3$ can be
rationalized qualitatively as follows. The bottom of the cylinders with
no ring offers small tangential friction to the sand surface during
the first moments of the fluidization process (when they are on top
of the granular surface), which implies a small torque between the
horizontal friction at the bottom and the horizontal inertial force
that can be represented at the center of mass of the cylinder. So,
the cylinder keeps its vertical position since the beginning of the
process, and just sinks vertically into the sand due to the action
of gravity.

Differently from No-ring cylinders, the basement of a Ring one is
firmly settled in the granular material, so, when an acceleration
is imposed by the shaker, a significant torque appears, arising
from the force  exerted by the sand on the ring, and the inertial
force at the cylinder's center of mass, forcing it to tilt. Moreover,
he presence of the ring prevents the free flow of sand near the bottom
of the cylinder, which makes more difficult its vertical sinking: as
the sink time increases, the torque caused by horizontal forces has
\textquotedblleft better possibilities\textquotedblright \ to tilt
the cylinder.

Two additional factors influence the sinking dynamics of this type
of intruder: firstly, the tilting process changes the effective size
of the cylinder, changing the drag and hydrostatic forces (in a way
that will be analyzed below), which contributes to a Brazil nut like
effect: isodense large particles tend to rise during shaking.
Additionally, when the inclination is high, most of the cylinder
could be inside the sand, and it starts to float, preventing further
tilting.

Comparing figures \ref{Fig7} (a) and (b) it is possible to see that
when the cylinder tilts it reaches a smaller final depth that when
it sinks without tilting, for similar adimensional accelerations. It
supports the idea that the inclination of the body increases the
resistance forces acting on it.

\begin{figure}
\includegraphics[width=0.4\textwidth]{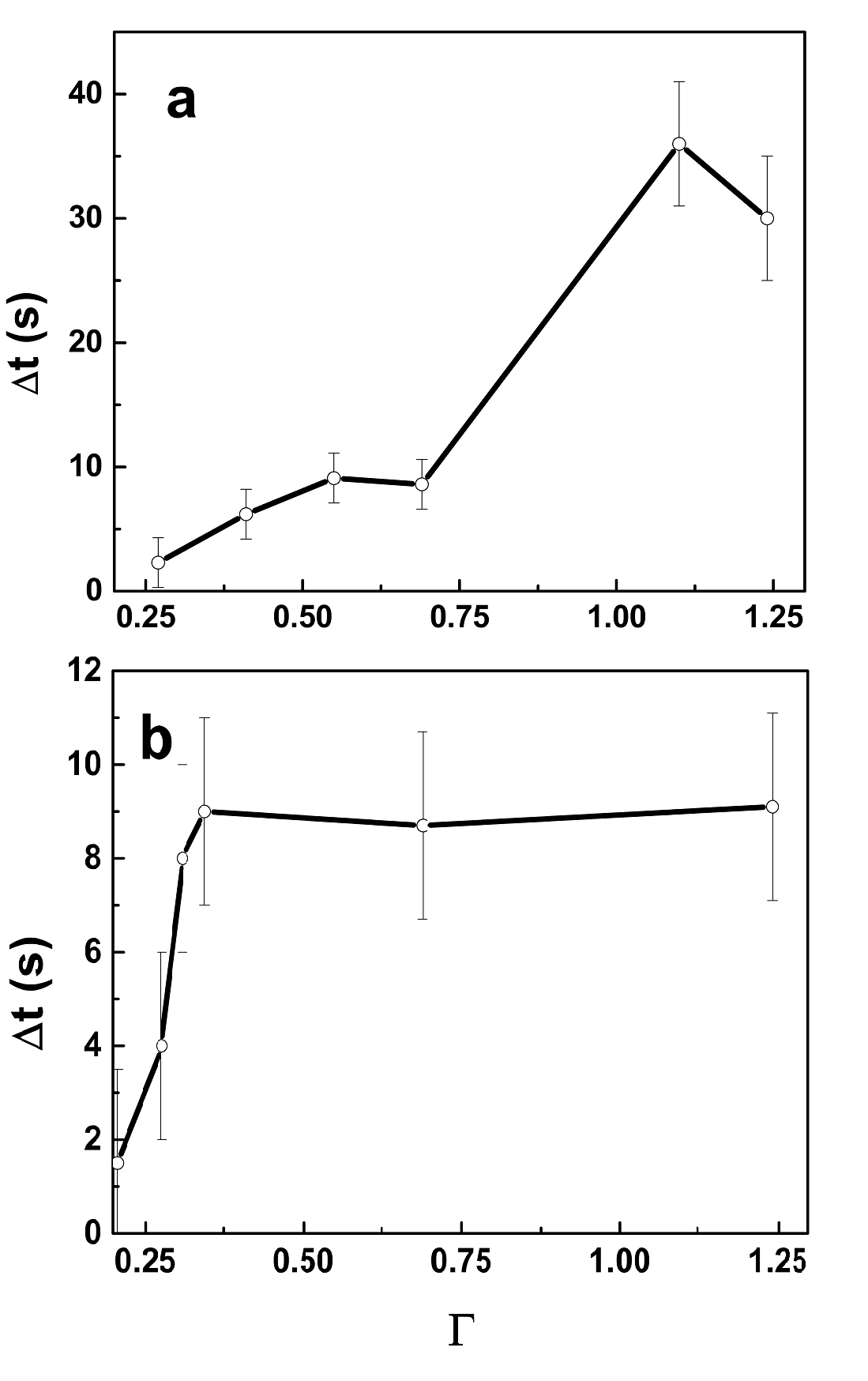}
\centering \caption{Sinking times for No-ring cylinders (a) and tilting times
for Ring cylinders (b). Sinking times for Ring cylinders are almost
equal to their tilting times, so they are not shown.}
\label{Fig8}
\end{figure}

Figure \ref{Fig8}(a) reports the time interval needed by No-ring cylinders to
penetrate into the granular material, from the moment when the
vibration was turned ON, until they find their new equilibrium
position. If we compare this figure with figure \ref{Fig7}(a),
a main difference can be appreciated: the plateau is now reached at a higher
acceleration. Anyway, both the sinking depth and sinking times increase
monotonically with the increase of $\Gamma$ and saturate for accelerations
at which the cylinder is eventually completely immersed in the material.

Figure \ref{Fig8}(b)  shows a different tendency: the Ring cylinders
tilt very fast, reaching a final inclination long before the final
depth of the No-ring ones is reached. A careful inspection of the
videos shows that No-ring cylinders sink fast during the first few
seconds, but dramatically slow down during the penetration of the
last millimeters, resulting in long sinking times. On the other
hand, when Ring cylinders end the tilting process, they do not
significantly sink further, perhaps due to the increase of the
friction and buoyancy. That is why we do not show the sinking times
of Ring cylinders: they are almost equal to the times shown in Fig.
\ref{Fig8}(b).

When we compare this  figure with Fig. \ref{Fig7}(c), an important
difference is easily appreciated: though the tilt angle increases
monotonically with $\Gamma$, the tilt times increase abruptly, and
then saturate. That behavior could be related with the fact that,
for small accelerations, the tilting process occurs slowly, but the
tilting angle is small. By contrast, for bigger accelerations the
tilting angle is larger, but the tilt occurs at bigger angular
velocities, giving almost constant tilting times ($\Delta t= \theta
/ \omega$).

\subsection{A phenomenological Newtonian model}

In order to formulate a model to describe analytically the sinking process,
let us consider the forces acting on the cylinder. As soon as the shaking
starts, if the adimensional acceleration is above the threshold, the upper
part of the granular bed is fluidized, and the intruder sinks.

Let us assume that the cylinder just sinks vertically, and let us
name the vertical downward axis as $z$. The force balance on the
intruder can be written as

\begin{equation}
 \label{eq.2}
 m \vec a = m \vec g + \int (-P)  \hat{n} dS +\int  \sigma_s \cdot \hat{n} dS
\end{equation}

\noindent where $P$ is the pressure, $\sigma_s$ the shear stress
tensor, $\hat{n}$ is the vector normal to the intruder's surface,
and the integrals run over the boundary of the intruder that is
inside the granular material. Assuming a hydrostatic pressure
profile, we can write:

\begin{equation}
 \label{eq.3}
  P= \int_0^h \rho(z') g dz'
\end{equation}

\noindent where $h$, as previously, is the depth reached by the
cylinder below the surface of the granular medium. In
Eq. (\ref{eq.3}) we have made explicit that the density of the
material varies with depth. Let us assume that it varies as a power
law between zero and the density of the solid layer, $\rho_{sl}$,
that is reached at a depth $h_f$:

\begin{equation}
 \label{eq.4}
 \rho(z')=\rho_{sl} \Big ( \dfrac {z'} {h_f} \Big )^p
\end{equation}

\noindent where $p \in [0,1]$.

By combining (\ref{eq.4}) and (\ref{eq.3}) and integrating, we find
the hydrostatic buoyancy force acting on the cylinder with a length
$h$ under the (average) level of the granular bed, as:

\begin{equation}
 \label{eq.5}
 \int (-  {P})  \hat{n} dS=- \dfrac {\rho_{sl} S g } {(p+1) h_f^p} h^{p+1} \hat h
\end{equation}

\noindent where $S$ is the characteristic area of the intruder
cross section, and $\hat h$ is a unit vector pointing downwards. It is
easy to see that the buoyancy force depends on the volume submerged
into the granular medium.

Assuming that inertial forces can be neglected, the shear stress component goes as

\begin{equation}
 \label{eq.6}
\int \sigma_s \cdot \hat n dS = -D \gamma v \hat h
\end{equation}

\noindent where $\gamma$  has the dimensions of a viscosity, $D$ is
the characteristic size of the cross section of the intruder, $v$ is
its sinking speed and $\hat h$ is the unit vector pointing downwards
\cite{deBruyn2004, Hou2005}. By substituting Eq. (\ref{eq.5}) and
Eq. (\ref{eq.6}) into Eq. (\ref{eq.2}), and only recovering the
modular values, we get:

\begin{equation}
 \label{eq.7}
m \dfrac {d^2h} {dt^2}+ D \gamma \dfrac {dh} {dt} + \dfrac {\rho_{cs} S g } {(p+1) h_f^p} h^{p+1} =mg
\end{equation}

Before solving Eq. (\ref{eq.7}) we will assume that the sink
velocity is constant, which follows quite well the behavior during
the fast sink regime, as seen in Fig. \ref{Fig8} ({\it i.e.}, we
neglect the inertial term). So,

\begin{equation}
 \label{eq.8}
 \dfrac {dh} {dt} + \dfrac {\rho_{sl} S g } {D \gamma (p+1) h_f^p} h^{p+1} =\dfrac {mg} {D \gamma}
\end{equation}

\noindent which can be written as

\begin{equation}
 \label{eq.9}
 \dfrac {dh} {dt} + a h^{p+1} =b
\end{equation}

The definitions of $a$ and $b$ are easily deduced comparing Eqs. (\ref{eq.8}) and
(\ref {eq.9}).

Equation (\ref{eq.9}) has analytical solutions if $p=0$ or $p=1$,
which correspond to the extreme cases of constant density and a
linear density profile with depth, respectively. The solutions are

 \begin{equation}
\label{eq.10}
h(t)=\dfrac b a (1-e^{-a t})
\end{equation}

\noindent if $p=0$, and

 \begin{equation}
\label{eq.11}
h(t)=\sqrt {\dfrac b a} \tanh (\sqrt {a b} t)
\end{equation}

\noindent if $p=1$.

It is easy to see that both expressions correspond to an exponential
growth that saturates.

\begin{figure}
\includegraphics[width=0.4\textwidth]{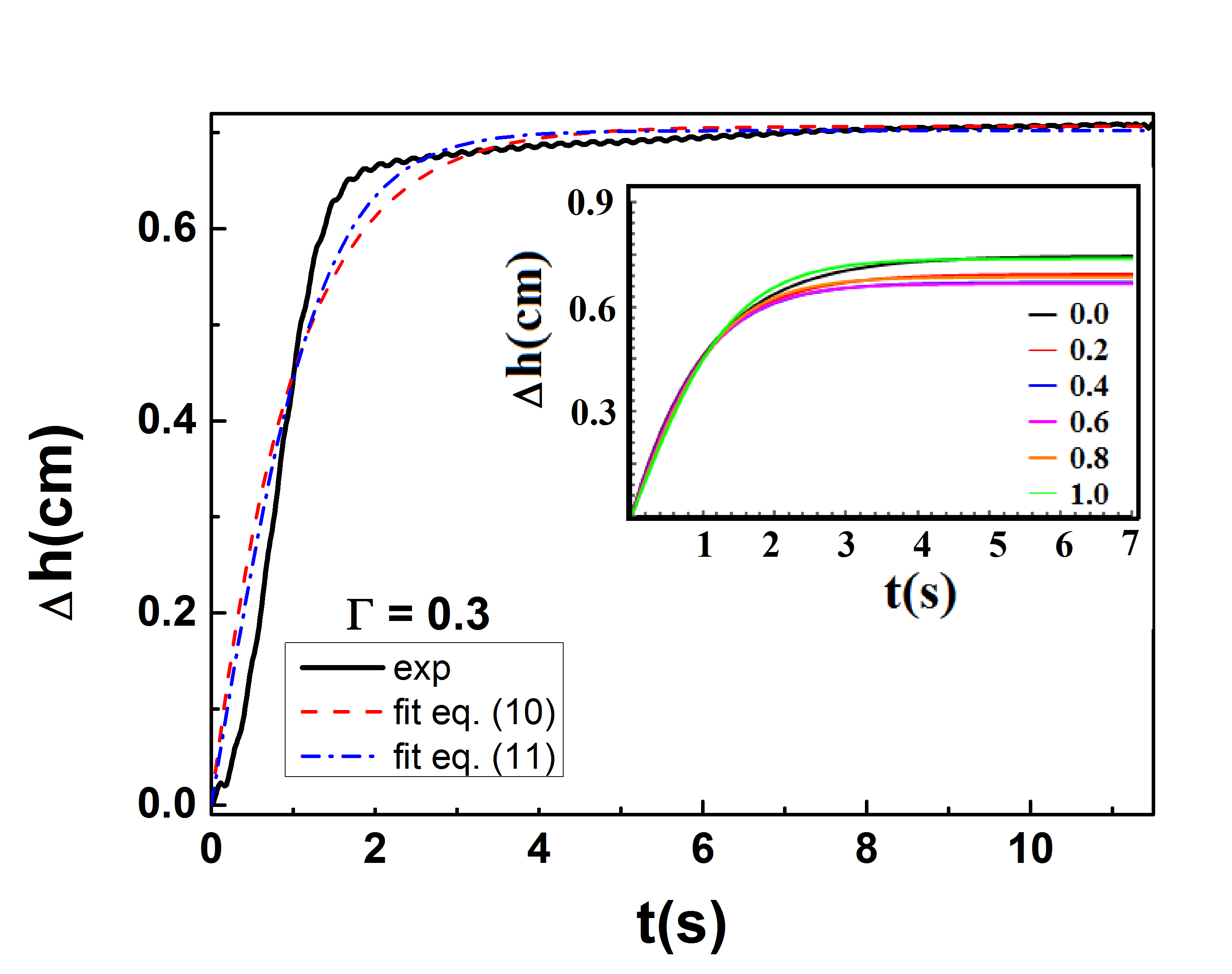}
\centering \caption{(color online) Time dependence of sinking depth
for the No-ring cylinder from experiment, compared with that
determined from Eqs. (\ref{eq.10}, \ref{eq.11}). The inset shows the
solutions of Eq. (\ref{eq.7}) for different values of $p$ (see
text). } \label{Fig9}
\end{figure}

Figure \ref{Fig9} shows the experimental results (continuous line)
obtained for an adimensional acceleration $\Gamma \simeq 0.3 $. It
is possible to see in more detail the initial fast sinking process,
followed by the slow creep.  Fig. \ref{Fig9} also shows the fitting
of equations (\ref{eq.10}, \ref{eq.11}) to experimental data. Even
the simplified Eq. (\ref{eq.10}) reproduces well the main features
of the sinking process.

It is almost impossible to determine experimentally the exact
density profile. But we do not need to know it in order to validate
our model, if we use the following rationale. Firstly, we fit Eqs.
(\ref{eq.10}, \ref{eq.11}) to the experimental data and obtain the
values of $a$, $b$ that correspond to $p=0$ ($a(p=0),b(p=0)$) and
$p=1$ ($a(p=1),b(p=1)$). Let us assume that $a$ and $b$ vary
linearly with $p$ between the extremes values which were obtained
from the fitting process. For an intermediate value of $p$ (say,
$p_1$) we can calculate the corresponding values of $a(p_1)$ and
$b(p_1)$. With them, we can in turn determine the constants of Eq.
(\ref{eq.7}). Then, we solve this equation numerically. This
procedure is repeated for values of $p$ between 0 and 1, with a step
of 0.1.

The inset in Fig. \ref{Fig9} shows some of the numerical solutions
for the values of $p$ in the legend. The main conclusion is that the
density profile has small influence on the first (and most
important) part of the sinking process. Of course, the final depth
is influenced by the value of $p$, but due to experimental
uncertainties, it is almost impossible to choose any particular
value.

Let us now study the influence of the values of $p$ in the quality
of the fit of the solution of Eq. (\ref{eq.7}) to the experimental
data. For doing this we note that the values of $a$ and $b$ in Eq.
(\ref{eq.10}) can be easily obtained from the experiments.
Considering Eq. (\ref{eq.9}) in the first moments of motion, as $h$
is small, $h'(t) \simeq b$, so $b$ can be evaluated as the initial
slope. As at large times $h(t)\sim h_{eq}$ ($h_{eq}=\Delta h $ if
$h(0)=0$) then $a=b/h_{eq}^{p+1}$. Then solving Eq. (\ref{eq.7}) for a
given value of $b, p, a(p)$ and naming the result $h_{mod}$, the
best value of $p$ arises from the minimization:

\begin{equation}
\label{eq.12}
 min \sum_{i=1}^N (h_{mod}(t,p)-h_{exp}(t))^2
\end{equation}

\noindent where $h_{exp}(t)$ are the experimental values of $h$.

The result for $\Gamma \leq 1.0$ is indifferent to $p$: the fit is
equally good no matter which is the value of $p \in [0,1]$. For
$\Gamma = 1.24$ there are differences in the quality of fits for
different $p$, but Eq. (\ref{eq.12}) gives a minimum for $p=0$,
so, we will assume $p=0$ in the following. Then, Eq. (\ref{eq.7})
becomes:

\begin{equation}
 \label{eq.13}
m \dfrac {d^2h} {dt^2}+ D \gamma \dfrac {dh} {dt} + \dfrac {\rho_{sl} S g } { h_f} h =mg
\end{equation}

\noindent that can be taken as the simplest equation of motion that
describes the vertical sink dynamics of our cylinders. It is worth
noticing that Eq. (\ref{eq.13}) reproduces quite closely the results
reported in Fig. \ref{Fig8}, and can be used to explain the vertical
sinking of Ring-cylinders while tilting, as we will see below.

Before that, it is instructive to compare equation (\ref{eq.13}) for
a granular bed fluidized by shaking, with that proposed in
\cite{Pacheco2011} to describe the penetration of an intruder into
ultra-light granular material that eventually behaves like a fluid
medium even in the absence of shaking. The equation proposed in
\cite{Pacheco2011} reads as:

\begin{equation}
 \label{eq.14}
 m \dfrac {d^2h} {dt^2}= mg - \eta \Big ( \dfrac {dh} {dt} \Big )^2 - \kappa \lambda \Big (1- e^{- {\dfrac {h} {\lambda}}} \Big)
\end{equation}

Firstly we will analyze the last term of Eq. (\ref{eq.14}) which is
related to Janssen's pressure. If  we were applying this equation to
our system, we should have taken into account that the shaking
promotes the destruction of the force chains. That is equivalent to
assume a very big $\lambda$, which yields a pressure force of the
form $\kappa \lambda (z /\lambda) = \kappa z$, as proposed earlier
in \cite{Katsuragi2007}. Comparing with Eq. (\ref{eq.13}) we see
that the depth-dependent terms in both equations are similar,
provided that

\begin{equation}
 \label{eq.15}
 \kappa= \dfrac {\rho_{sl} S} {h_f} g
\end{equation}

\noindent (notice that the linear relation of $\kappa$ with $g$ has
been demonstrated experimentally in \cite{Altshuler2014}).

However, there is an important difference in the \textquotedblleft viscous drag\textquotedblright
terms between equations (\ref{eq.13}) and (\ref{eq.14}): in the
first, the velocity is linear, while it is squared in the second.
Indeed, we cannot reproduce our experimental sink process if we
insert in Eq. (\ref{eq.13}) a squared velocity term. However, the
difference can be justified by the fact that the sink velocities in
our shaken-bed experiment are much smaller than those observed in
the penetration experiments reported in \cite{Pacheco2011} and
\cite{Altshuler2014}.

Now we concentrate again in the interpretation of our experimental
results. In order to understand the tilting dynamics, it is useful
to note that, when applying Eq. (\ref{eq.13}) to a tilted cylinder,
the values of both $D$ and $S$ change. The reason is that when we
calculate the surface integral, the result will be proportional to
the cylinder's immersed volume. As the cylinder tilts, the immersed
surface increases more than in the case of sinking without tilting,
so the drag force is bigger in the former case. Considering, for
instance, the situation represented in the lower row of  Fig.
\ref{Fig3}, when the cylinder sinks a distance  $\Delta h $, the
surface increases as the inverse of $\cos \theta$ (of course,
other intruder geometries may follow different laws).

To test it, let us assume a simplified model: the increase factor of
$S$ and $D$ is proportional to the characteristic size of the cross
section of the cylinder projected on the horizontal plane, i.e., it
is proportional to the inverse of $\cos \theta$. Then, instead of
$D$ and $S$, we will solve Eq. (\ref{eq.13}) using $D / \cos \theta (t)$
and $S / \cos \theta(t)$, where $\theta (t)$ is a function that grows
from zero to the maximum angle $\theta_{max}$ reached by the cylinder,
mimicking Fig. \ref{Fig6}(c).

\begin{figure}
\includegraphics[width=0.4\textwidth]{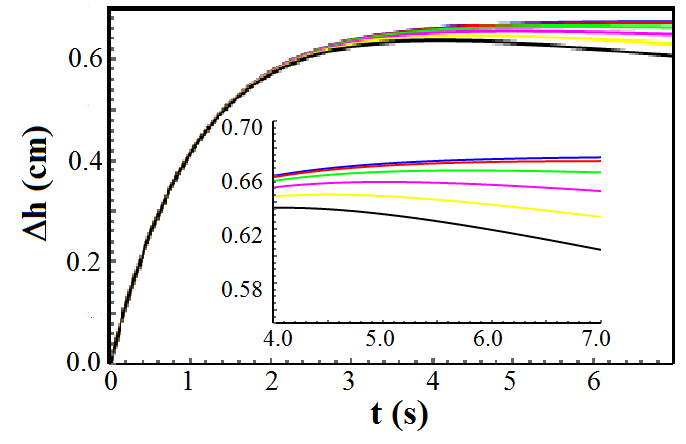}
\centering \caption{(color online) Time dependence of sinking depth
as calculated solving numerically Eq. (\ref{eq.13}) considering the variation of
$S$ and $D$ provoked by tilting (see text). Upper
curve is for $\theta_{max}=$ 0 while the lower one is for $\theta_{max}= \pi /
3$. Between them, $\theta_{max}$ varies in steps of $\pi/15$. The inset shows
the last three seconds.
}
\label{Fig10}
\end{figure}

The consequences can be seen in Fig. \ref{Fig10}. While in the beginning the sinking
process in all situations occurs with the same dynamics, as the cylinder approaches
the final angle, the behavior changes, been the final depth larger for the situations
corresponding to low  tilting.

The upper curve, calculated for $\theta=$0 coincides with the upper
curve in the inset of Fig. \ref{Fig9} (calculated for $p=0$). Subsequent curves are
calculated for values of $\theta_{max}$ varying in steps of $\pi /$15, the
lowermost curve corresponds to $\theta_{max}=\pi/$3. As the inclination of
the cylinder increases, both the buoyancy and the viscous drag do.
The effect of these factors on the sinking process of Ring cylinders
was already noted in Fig. \ref{Fig6}(b): an immediate consequence is
the decrease of the sinking depth (for a given $\Gamma$), compared with
that of the No-ring ones, which can be easily observed in
the experiments. In the inset it is possible to deduce that, for the
larger angles, a small decrease in the depth is observed, suggesting
the influence of a Brazil nuts like effect.

In spite of the simplifications assumed, it is worth noting that the
basic differences between Fig. \ref{Fig7} (a) and (b) are reproduced
by our model.

Finally, there is another element that was also not considered
in our model: as the container shakes horizontally, it is equivalent to a
horizontal drag that changes periodically its direction. According
to \cite{Zhang2015}, it creates an additional lift force, and also a
dependence of the drag force with depth, which, of course, must
influence the detailed penetration dynamics of the Ring cylinders.

\subsection{Quasi-2D numerical simulations}

\begin{figure}
\includegraphics[width=0.4\textwidth]{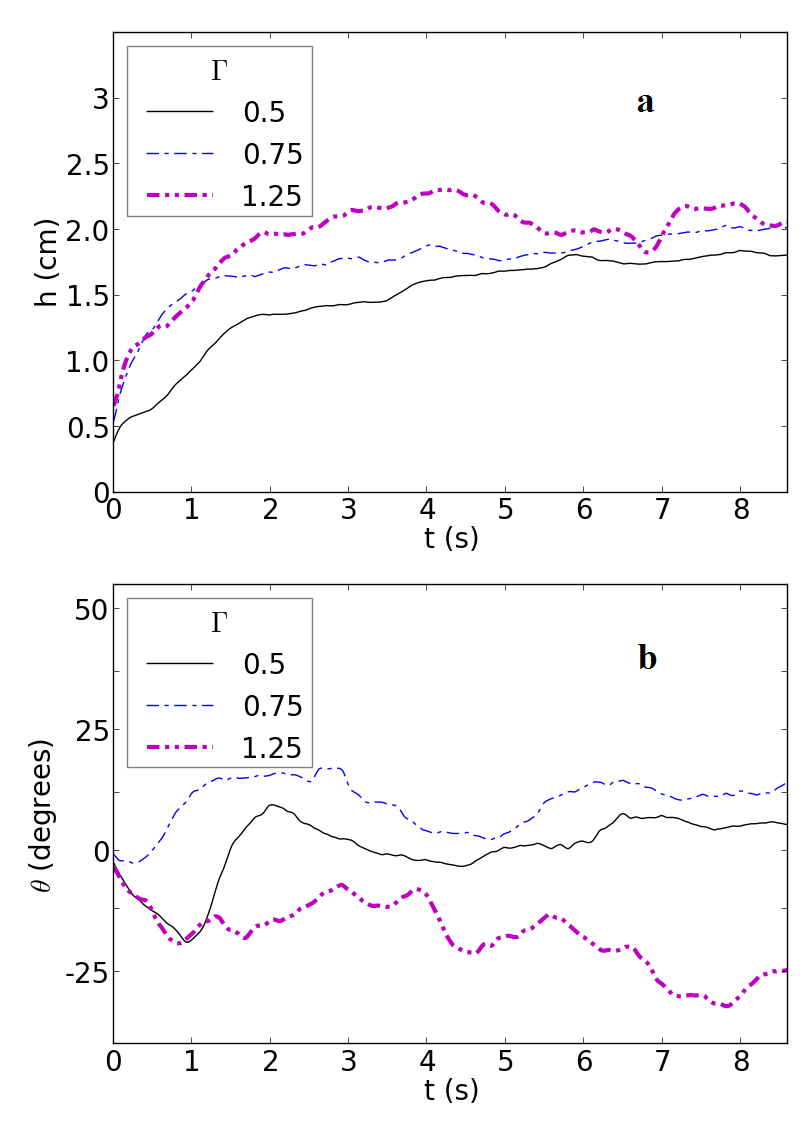}
\centering \caption{(color online) Time dependence of sinking depth (a)
and tilting angle (b) for a simulated quasi-2D No-ring cylinder.
}
\label{Fig11}
\end{figure}

\begin{figure}
\includegraphics[width=0.4\textwidth]{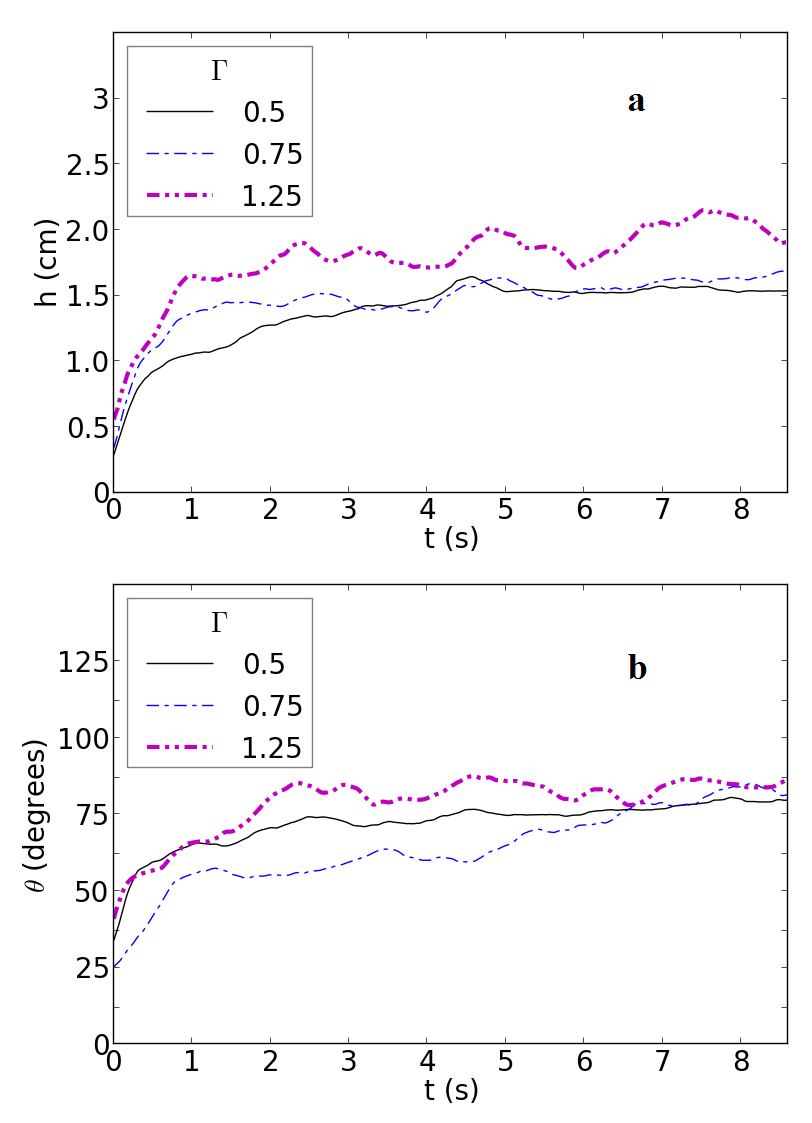}
\centering \caption{(color online) Time dependence of sinking depth (a)
and tilting angle (b) for a simulated quasi-2D Ring cylinder. }
\label{Fig12}
\end{figure}

In figure \ref{Fig11} we can see the behavior of the quasi-2D
No-ring intruder for three different adimensional accelerations. The
same accelerations have been applied to a medium containing the
simulated Ring intruder, see Fig. \ref{Fig12}.

Regarding the vertical sinking, we do not observe major changes
between Ring and No-ring intruders. We can notice that they sink
less than in the experiments, which may be related with the
dimensionality of the simulation relative to the real experiment.
Quasi-2D granular media allow less choices of readjustment than in
3D: they are easily jammed, which makes it more difficult for an
object to sink. Moreover, the size ratio of the intruder over the
particles is 15 times smaller in the simulations than in the
experiments (experiments: 44mm/0.280mm = 150; simulations: 40mm/4mm
= 10), which means that if one particle is stuck under the intruder
during the simulations, it will slow down the intruder significantly
more than if the particle were 15 times smaller. In addition, since
the numerical intruder is made out of particles of the same size as
the particles of the medium, it may be rougher that the experimental
intruder.

For the tilting, we find that the presence of foundation at the
bottom of the intruder causes a large tilting. Indeed, for the
shaking with No-ring, the intruder tilting angle is a few tens of
degrees, but during the shaking with ring, the intruder tilts fast,
reaching a angle between 80 and 90\textdegree  (see Fig.
\ref{Fig5}): the intruder almost ends up lying on one of its
sides.

An interesting result of the simulations is seen in figure
\ref{Fig11} (b): while for the smaller accelerations the tilting
angle of the No-ring cylinders oscillate around 0\textdegree, for
the adimensional acceleration $\Gamma = 1.25$ the No-ring cylinder
reaches an angle of approximately 25\textdegree. It suggests that,
for big enough accelerations, it is possible to tilt the cylinder,
independently of the details of its foundation. Some preliminary
experimental results obtained by us using higher frequencies of
shaking (which gives higher adimensional accelerations) does confirm
that finding.

\subsection{Conclusions}

In this paper, we have studied the behavior of cylindrical objects
as they sink into a dry granular bed fluidized by horizontal
oscillations, as a model system to understand the effects of
earthquake-related fluidization of soils on human constructions and
other objects like rocks.

We have found that, within a relatively large range of lateral
shaking amplitudes at a frequency of 5 Hz, cylinders with flat
bottoms sink vertically, while those with a \textquotedblleft
foundation\textquotedblright consisting in a shallow ring attached
to their bottom, tilt laterally besides vertically sink. The tilting
is associated to the torque experienced by the cylinder when the
ring at the bottom increases the friction with the
laterally-accelerated granular bed.

We have been able to mimic the above described behaviors by quasi-2D
numerical simulations. We have also reproduced the vertical sink
dynamics of cylinders with a flat base using a Newtonian equation of
motion for an object penetrating a fluidized layer of granular
matter, where the granular effective density increases with depth,
eventually reaching a solid phase. The same model allows to
understand the sinking even in the present of tilting.

Finally, it is worth noting that preliminary experimental data and
quasi-2D numerical simulations suggest that, when strong enough
lateral shaking is applied, the tilting scenario tends to dominate
regardless the nature of the intruder's foundation.

\section {Acknowledgements}
 We acknowledge support from Project 29942WL (Fonds
de Solidarit\'e Prioritaire France-Cuba), from the EU ITN
FlowTrans, and from the Alsatian network REALISE. E. A. drew
inspiration from the late M. \'Alvarez-Ponte.

\end{document}